\documentclass{emulateapj}

\usepackage{amssymb}
\usepackage{amsmath}
\usepackage{graphicx}
\usepackage{natbib}

\usepackage{txfonts}

\usepackage[breaklinks,colorlinks=true,linkcolor=red,citecolor=blue]{hyperref}

\shorttitle{On the magnetic field of pulsars with realistic neutron star configurations}
\shortauthors{R.~Belvedere, J.~A.~Rueda, R.~Ruffini}

\begin{document}
\title{On the magnetic field of pulsars with realistic neutron star configurations}
\author{R.~Belvedere, Jorge A. Rueda, R.~Ruffini}
\affil{Dipartimento di Fisica and ICRA, Sapienza Universita' di Roma P.le Aldo Moro 5, I-00185 Rome, Italy}
\affil{ICRANet, P.zza della Repubblica 10, I-65122 Pescara, Italy}
\affil{ICRANet-Rio, Centro Brasileiro de Pesquisas F\'isicas, Rua Dr. Xavier Sigaud 150, Rio de Janeiro, 22290--180, Brazil}

%
%


\altaffiltext{}{riccardo.belvedere@icra.it, jorge.rueda@icra.it, ruffini@icra.it}

\begin{abstract}
We have recently developed a neutron star model fulfilling global and not local charge neutrality, both in the static and in the uniformly rotating cases. The model is described by the coupled Einstein-Maxwell-Thomas-Fermi (EMTF) equations, in which all fundamental interactions are accounted for in the framework of general relativity and relativistic mean field theory. Uniform rotation is introduced following the Hartle's formalism. We show that the use of realistic parameters of rotating neutron stars obtained from numerical integration of the self-consistent axisymmetric general relativistic equations of equilibrium leads to values of the magnetic field and radiation efficiency of pulsars very different from estimates based on fiducial parameters assuming a neutron star mass, $M=1.4~M_\odot$, radius $R=10$~km, and moment of inertia, $I=10^{45}$~g~cm$^2$. In addition, we compare and contrast the magnetic field inferred from the traditional Newtonian rotating magnetic dipole model with respect to the one obtained from its general relativistic analog which takes into due account the effect of the finite size of the source. We apply these considerations to the specific high-magnetic field pulsars class and show that, indeed, all these sources can be described as canonical pulsars driven by the rotational energy of the neutron star, and with magnetic fields lower than the quantum critical field for any value of the neutron star mass.
\end{abstract}

\keywords{Neutron Stars; Nuclear Matter Equation of State; General Relativity.}

\section{Introduction}\label{sec:1}

The traditional formula used in pulsar astrophysics literature to infer the pulsar's magnetic field originated from the pioneer hypothesis of \citet{1968Natur.218..731G} and \citet{1968Natur.219..145P}, who first stressed the role of the rotational energy of the neutron star as an energy reservoir for the pulsar's activity. The surface magnetic field of pulsars has been since then estimated \citep[see, e.g.,][]{1969Natur.221..454G,1969ApJ...157.1395O,ferrari69} by equating the rotation energy loss of the neutron star,

\begin{equation}\label{eq:Edot}
\dot{E}_{\rm rot} = - 4 \pi^2 I \frac{\dot{P}}{P^3}\, ,
\end{equation}

to the radiating power of a rotating magnetic point dipole in vacuum,

\begin{equation}\label{eq:dipole}
P_{\rm dip}=-\frac{2}{3}\frac{\mu^2_\bot \Omega^4}{c^3}\, .
\end{equation}

Here $\Omega$ is the rotation angular velocity of the star, $\mu_\bot=\mu\sin\chi$ is the component of the magnetic dipole $\mu=B R^3$ perpendicular to the rotation axis being $B$ the magnetic field at the equator, and $\chi$ denotes the inclination angle of the magnetic dipole with respect to the rotation axis. Under these assumptions, the magnetic field is estimated as

\begin{equation}\label{eq:Bmax}
B\sin\chi=\left(\frac{3 c^3}{8 \pi^2} \frac{I}{R^6} P \dot{P} \right)^{1/2}\, ,
\end{equation}

where $P=2\pi/\Omega$ and $\dot{P}$ are the rotational period and the spin-down rate of the pulsar, which are observational properties, while the moment of inertia $I$ and the radius $R$ of the star are model dependent properties.

It is worth to recall that the electromagnetic power given by Eq.~(\ref{eq:dipole}) of the above simplified rotating magnetic point dipole model coincides, in the so-called \emph{wave zone} approximation ($r\gg c/\Omega=1/k=\lambda/2\pi$, where $k$ is the wave number and $\lambda$ the wavelength), with the one obtained from the classic work by \citet{1955AnAp...18....1D}, where the exterior (vacuum) electromagnetic field of a uniformly rotating, perfectly conducting star with a misaligned magnetic dipole were obtained as an exact closed-form analytic solution of the Maxwell equations in flat spacetime.

General considerations on the nature of pulsars are often extracted in the literature from the application of the above formulas with fiducial parameters of a pulsar: a canonical neutron star of mass $M=1.4 M_\odot$, radius $R=10$ km, and moment of inertia $I=10^{45}$ g cm$^2$, \citep[see e.g.][and references therein]{caraveo14}. For these fiducial parameters, Eq.~(\ref{eq:Edot}) becomes
\begin{equation}\label{eq:Edotf}
\dot{E}^f_{\rm rot} = -3.95\times 10^{46} \frac{\dot{P}}{P^3}\quad {\rm erg~s}^{-1}\, ,
\end{equation}
and, Eq.~(\ref{eq:Bmax}),
\begin{equation}\label{eq:BmaxNS}
B_f\sin\chi=3.2\times 10^{19} \left(P \dot{P} \right)^{1/2} {\rm G}\, .
\end{equation}

We focus in this work in the interesting class referred to as high-magnetic field pulsars \citep[see, e.g.,][]{kaspi11}. In Table \ref{tab:H-B_pulsar} \citep{kaspi11,zhu11}, we show a sample of the high-magnetic field pulsar class with their properties as inferred from the fiducial formulas (\ref{eq:BmaxNS}) and (\ref{eq:Edotf}) for the surface magnetic field and rotational energy loss (see second and fourth column of the table, respectively). Notice that magnetic fields with values higher than the critical field for quantum electrodynamical effects,
\begin{equation}\label{eq:Bc}
B_c=\frac{m^2_e c^2}{e \hbar}=4.41\times 10^{13}~{\rm G}\, ,
\end{equation}
appear, and in some cases also luminosities higher than the rotational power of the neutron star, namely $L_X>|\dot{E}_{\rm rot}^f|$.

\begin{center}
\begin{deluxetable*}{lccccc}[!hbtp]
\tabletypesize{\scriptsize}
\tablecaption{Properties of the high-magnetic field pulsars obtained assuming fiducial neutron star parameters, $R=10$~km and $I=10^{45}$~g~cm$^2$, respectively, and using Eq.~(\ref{eq:BmaxNS}) with inclination angle $\chi=\pi/2$ and Eq.~(\ref{eq:Edotf}). See \citet{zhu11,kaspi11} for additional details of these pulsars.\label{tab:H-B_pulsar}}
\tablewidth{0pt}
\tablehead{
\colhead{Pulsar} & \colhead{$B_f/B_c$} & $L_X$~($10^{33}$~erg~s$^{-1}$) & $L_X/|\dot{E}_{\rm rot}^f|$                & $P$~(s) & $\dot{P}$~($10^{-12}$)
}
\startdata
J1846--0258      & 1.11                & $25-28^a$, $120-170^b$         & $0.0031-0.0035^a$, $0.015-0.021^b$ & 0.326   & 7.083              \\
J1819--1458$^c$  & 1.13                & $1.8-2.4$                      & $6.21-8.28$                        & 4.263   & 0.575              \\
J1734--3333      & 1.18                & $0.1-3.4$                      & $0.0018-0.0607$                    & 1.169   & 2.279              \\
J1814--1744      & 1.24                & $<43$                          & $<91.5$                            & 1.169   & 2.279              \\
J1718--3718      & 1.67                & $0.14-2.6$                     & $0.0875-1.625$                     & 3.378   & 1.598              \\
J1847--0130      & 2.13                & $<34$                          & $<200$                             & 6.707   & 1.275              
\enddata
\tablenotetext{a}{in 2000, prior to the 2006 outburst}
\tablenotetext{b}{during the outburst in 2006}
\tablenotetext{c}{classified as a rotating radio transient (RRAT)}
\end{deluxetable*}
\end{center}

Due to these theoretically inferred properties, it has been suggested the possibility that this family of pulsars can be the missing link, i.e. transition objects, between rotation-powered pulsars and the so-called magnetars: neutron stars powered by the decay of overcritical magnetic fields. In principle this would lead to a large unseen population of magnetars in a quiescence state which could be disguised as radio pulsars \citep[see, e.g.,][]{zhu11}.

However, as we shall show in this work, these conclusions might be premature since the surface magnetic fields inferred by fiducial neutron star parameters, namely by Eq.~(\ref{eq:BmaxNS}), are in general overestimated. Indeed, much lower values of the magnetic field are obtained when realistic structure parameters are applied and when general relativistic corrections are introduced to traditional Newtonian equation (\ref{eq:Bmax}); see section~\ref{sec:3}. The need of using more realistic neutron star configurations is the result of the knowledge of a more complex nuclear equation of state, structure, and stability conditions of both static and rotating neutron stars, acquired in the intervening years from the seminal work of \cite{oppenheimer39}.

We show the results for neutron stars in two cases of interest: 1) configurations obtained under the traditional constraint of local charge neutrality, and 2) configurations subjected to the constraint of global charge neutrality, in which the Coulomb interactions are introduced. For the latter configurations we use our recent formulation of the neutron star theory for both static and uniform rotation, following our previous works \citep{belvedere12,belvedere14}. These new set of equations, which we called Einstein-Maxwell-Thomas-Fermi equations, accounts for the weak, strong, gravitational and electromagnetic interactions within the framework of general relativity and relativistic nuclear mean-field theory. 

We shall show that independently on the theoretical model, different structure parameters as a function of the central density and/or rotation frequency of the star give rise to quite different quantitative estimates of the astrophysical observables with respect to the use of fiducial parameters. 

This work is organized as follows. In section~\ref{sec:2} we briefly summarize the equations of equilibrium and resulting structure from their integration of both static and uniformly rotating neutron stars. We analyze in section~\ref{sec:3} the estimates of the magnetic field and radiation efficiency of the high-magnetic field pulsars class. We summarize our conclusions in section \ref{sec:4}.

We use cgs units throughout the article unless otherwise specified.

\section{Neutron star structure}\label{sec:2}

We have recently shown \citep{rotondo11d,rueda11,belvedere12}, that in the case of both static and rotating neutron stars, the Tolman-Oppenheimer-Volkoff (TOV) system of equations \citep{oppenheimer39,tolman39} is superseded by the Einstein-Maxwell system of equations coupled to the general relativistic Thomas-Fermi equations of equilibrium, giving raise to the what we have called the Einstein-Maxwell-Thomas-Fermi (EMTF) equations. These new equations account for the weak, strong, gravitational and electromagnetic interactions within the framework of general relativity and relativistic nuclear mean field theory. 

In the TOV-like approach the condition of local charge neutrality is applied to each point of the configuration, while in the EMTF equations the condition of global charge neutrality, is imposed. It was shown in \citep{rotondo11d,rueda11} that the approach based on local charge neutrality is inconsistent with the equations of motion of the particles in the system. Consequently, the general relativistic thermodynamic equilibrium of the star, first introduced by \cite{klein49} in the case of a self-gravitating one-component system of uncharged particles, is not satisfied when local charge neutrality is applied to a multi-component system with charged constituents. The equilibrium is ensured by the constancy, along the whole configuration, of the generalized electro-chemical particle potentials for all the species, what we denominated to as conservation of ``Klein potentials''. When finite temperatures are considered, the constancy of the gravitationally redshifted temperature \citep{tolman30} has to be also imposed \citep{rueda11}. 

The weak interactions are introduced via the condition of $\beta$-equilibrium. For the strong interactions we follow the $\sigma$-$\omega$-$\rho$ nuclear model within relativistic mean field theory \'a la \cite{boguta77b}.  The nuclear model is fixed by the coupling constants and the masses of the three mesons. We here adopt the NL3 parameter set \citep{lalazissis97}: $m_\sigma$=$508.194$ MeV, $m_\omega$=$782.501$ MeV, $m_\rho$=$763.000$ MeV, $g_\sigma$=$10.2170$, $g_\omega$=$12.8680$, $g_\rho$=$4.4740$, plus two constants that give the strength of the self-scalar interactions, $g_2=-10.4310$ fm$^{-1}$ and $g_3=-28.8850$.

The structure of the neutron star solution of the EMTF equations of equilibrium leads to a new structure of the neutron stars markedly different from the traditional configurations obtained through the TOV equations \citep[see Fig.~4 in][]{belvedere12}: from the supranuclear central density up to the nuclear density $\rho_{\rm nuc}\approx 2.7\times 10^{14}$~g~cm$^{-3}$, we find the neutron star core, which is composed by a degenerate gas of neutrons, protons, and electrons in $\beta$-equilibrium, and is positively charged.  The core is surrounded by an electron layer a few hundreds fermi thick that fully screens its charge. In this core-crust transition layer the electric field reaches values as large as $E\sim (m_\pi/m_e)^2 E_c$, where $E_c=m^2_e c^3/(e \hbar)\approx 1.3\times 10^{16}$~Volt~cm$^{-1}$ is the critical field for vacuum polarization. The $e^+e^-$ pair creation is however inhibited by Pauli blocking \citep{ruffinirep10}. In this layer the particle densities decrease until the point where global charge neutrality is reached and the crust is found. Consequently, the core is matched to the crust via this interface at a density $\rho_{\rm crust}\leq \rho_{\rm nuc}$. In the limit $\rho_{\rm crust}\to\rho_{\rm nuc}$, the thickness of the transition layer as well as the electric field inside it vanish, and the solution approaches the one given by local charge neutrality (see Figs.~3 and 5 in \cite{belvedere12}). The crust in its outer region $\rho \leq \rho_{\rm drip}\approx 4.3\times 10^{11}$ g~cm$^{-3}$ is composed by white dwarf-like material (ions and electrons), following for instance the BPS EOS \citep{baym71b}. In its inner region, at densities $\rho>\rho_{\rm drip}$, free neutrons are present and the EOS follows for instance the BBP description \citep{baym71a}. Configurations with $\rho_{\rm crust}>\rho_{\rm drip}$ possess both inner and outer crust while in the cases with $\rho_{\rm crust}\leq \rho_{\rm drip}$ the neutron star have only outer crust. As shown by \citet{belvedere12}, all the above new features lead to a new mass-radius relation of static neutron stars.

The extension of the above formulation to the case of uniform rotation has been recently achieved in \citep{belvedere14} within the Hartle formalism \citep{hartle67a}. It is worth to underline that the influence of the induced magnetic field owing to the rotation of the charged
core of the neutron star in the globally neutral case is negligible as we will show in subsection~\ref{subsec:2.1}. From the integration of the equations of equilibrium, we computed in \citet{belvedere14}, for different central densities $\rho_c$ and circular angular velocities $\Omega$, the mass $M$, polar $R_p$ and equatorial $R_{\rm eq}$ radii, angular momentum $J$, eccentricity $\epsilon$, moment of inertia $I$, as well as quadrupole moment $Q$ of the configurations. 

The angular momentum $J$ of the star is given by
\begin{equation}\label{eq:J}
J = \frac{1}{6}\frac{c^2}{G} R^4\left(\frac{d\bar{\omega}}{dr}\right)_{r=R}\;,
\end{equation}
which is related to the angular velocity $\Omega$ by
\begin{equation}\label{eq:Jomega}
\Omega = \bar{\omega}(R)+\frac{2 G^2}{c^5}\frac{J}{R^3}\;,
\end{equation}
where $R$ is the total radius of the non-rotating star and $\bar{\omega}(r)=\Omega-\omega(r)$ is the angular velocity of the fluid relative to the local inertial frame, with $\omega$ the fluid angular velocity in the local inertial frame. 

The total mass of the configuration is
\begin{equation}\label{eq:Mrot}
M = M_0+\delta M\;,\qquad \delta M = m_0(R)+\frac{G^2}{c^7}\frac{J^2}{R^3}\;,
\end{equation}
where $M_0$ is the mass of the non-rotating star and $\delta M$ is the contribution to the mass due to the rotation, while $m_0$ is a second order contribution to the mass related to the pressure perturbation.

\begin{figure}[!htbp]
\plotone{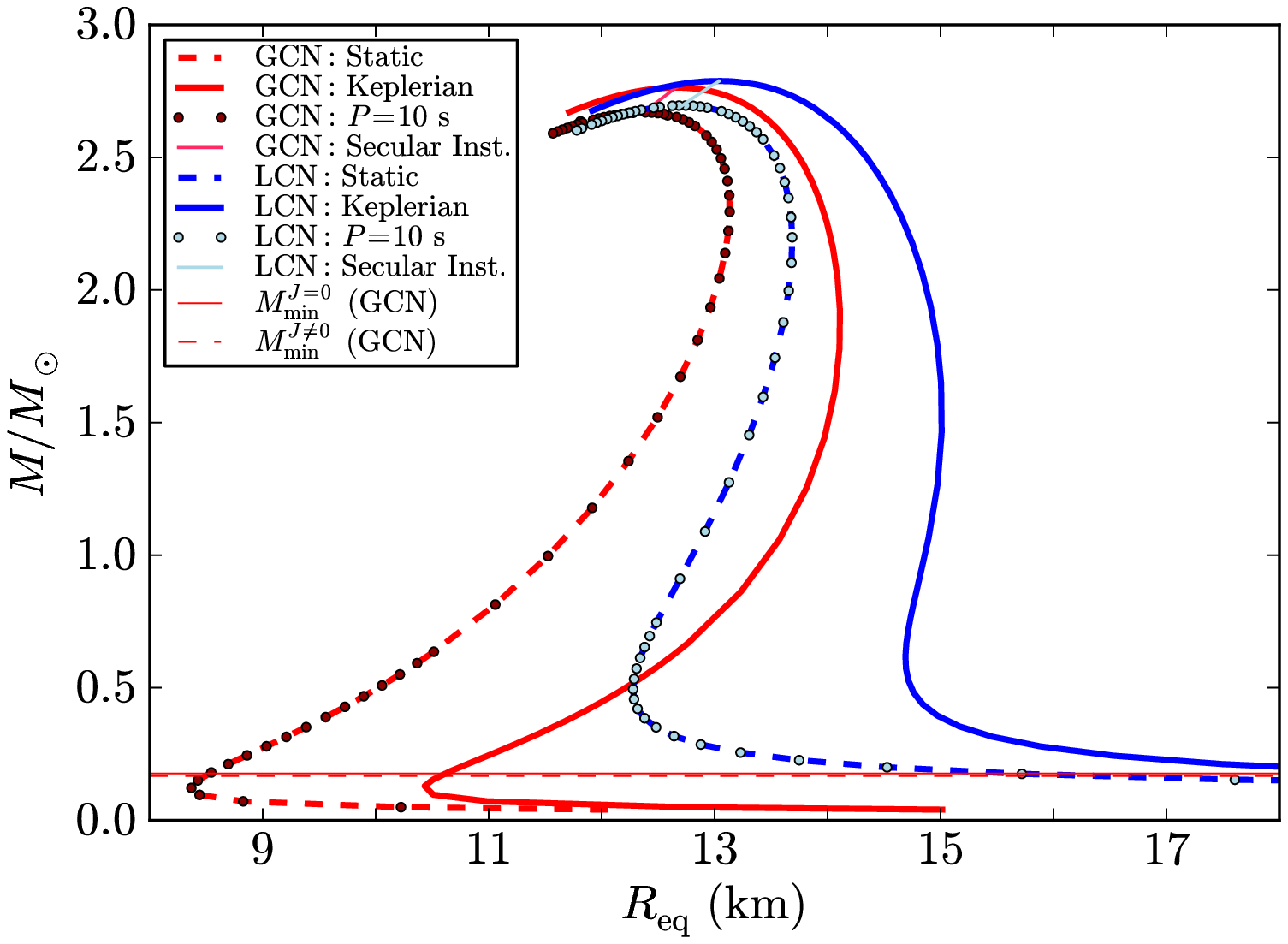}
\caption{Total mass versus total equatorial radius for the global (red) and local (blue) charge neutrality cases. The dashed curves represent the static configurations, while the solid lines are the uniformly rotating neutron stars. The pink-red and light-blue color lines define the secular instability boundary for the globally and locally neutral cases respectively. The horizontal thin red lines define the minimum mass in the globally neutral case. The dots refer to the sequence of constant period $P=10$~s.}\label{fig:MtotvsRtot}
\end{figure}

The moment of inertia can be computed from the relation
\begin{equation}\label{eq:I}
I = \frac{J}{\Omega}\;,
\end{equation}
which does not account for deviations from spherical symmetry since within the Hartle formalism $J$ is a first order function of $\Omega$. This is in any way a good approximation since, owing to the high density of neutron stars, most of the observed pulsars are accurately described by a perturbed spherical geometry. This can be seen for instance from the sequence of configurations with period $P=10$~s, shown in Fig.~\ref{fig:MtotvsRtot}, which practically overlaps the non-rotating mass-radius relation. The accuracy of the approximation increases for stiffer EOS \citep[see][for details]{benhar05}, as the ones given by $\sigma$-$\omega$-$\rho$ relativistic nuclear mean field models.

In Fig.~\ref{fig:MtotvsRtot} we show the mass-radius relation that results from the integration of the EMTF equations for the equilibrium configurations of static and rotating neutron stars. The dashed lines represent the non-rotating, ($J=0$), sequences, while the solid lines represent the corresponding maximally rotating (Keplerian) sequences. The pink-red and light blue lines represent the secular instability boundaries for the global and local charge neutrality cases, respectively. The horizontal thin red lines give the minimum mass for the static (solid line) and rotating (dashed line) sequences for the global charge neutrality case. This minimum mass limits are the configurations for which the gravitational binding energy vanishes, namely below this mass the neutron star is unbound. In the case of local charge neutrality case no minimum mass was found \citep[see][for further details]{belvedere14}. 

\subsection{Influence of the rotationally induced magnetic field}\label{subsec:2.1}

The interior electric field generates a magnetic field inside the neutron star once it is put into rotation. For the sake of clarity and without loss of generality, we now give an estimate of such an interior magnetic field by solving the Maxwell equations in flat Minkowski background. The charge distribution in the core and in the core-crust transition layer therefore rotate with constant angular velocity $\Omega$ around the axis of symmetry. The magnetic field can be first written in terms of the electromagnetic potential $\vec{A}$ as usual, i.e. $\vec{B}(\vec{r})=\vec{\nabla}\times\vec{A}(\vec{r})$. The electromagnetic potential can then be rewritten in terms of a new potential $\vec{F}(\vec{r})$ by $\vec{A}(\vec{r})=\vec{\Omega}/c^2\times\vec{F}(\vec{r})$, with $\vec{F}(\vec{r})=(4\pi\epsilon_0)^{-1}\int{(\vec{r'}\rho_{\rm ch}(\vec{r'})/(\vec{r}-\vec{r'}))d^3\vec{r'}}$. For a spherically symmetric charge distribution $\rho_{\rm ch}(r)$, the potential $\vec{F}$ can be taken as radial, i.e. $\vec{F}(\vec{r})=\vec{e_r}F(r)$, being $\vec{e_r}$ the unit radial vector \citep{marsh82}. The magnetic field is thus given by
\begin{equation}\label{eq:Bgeneral}
\vec{B}(\vec{r})=B_r\vec{e_r}+B_{\theta}\vec{e_{\theta}}\, ,
\end{equation}
where
\begin{equation}\label{eq:magcomponent}
B_r=\frac{2\Omega F}{c^2 r}\cos\theta \;,\qquad B_{\theta}=-\frac{2\Omega}{c^2}\left[\frac{F}{r}+\frac{r}{2}\frac{d}{dr}\left(\frac{F}{r}\right)\right]\sin\theta\, ,
\end{equation}
are respectively the radial and the angular component of the magnetic field, being $\theta$ the angle between $r$ and $z$ axis, and $\vec{e_{\theta}}$ the unit vector along $\theta$.

The Eqs.~(\ref{eq:Bgeneral}--\ref{eq:magcomponent}) can now be used to calculate the induced magnetic field both in the core and the core-crust interface shell surrounding it. Following \citet{boshkayev12a}, in order to estimate the rotationally induced magnetic field, we describe the core and the core-crust interface using a simplified model based on the previous works by \citet{rotondo11a,rotondo11b}. The distribution of $N_p$ protons, $n_p$, is assumed as constant within the core radius $R_c=\Delta \hbar/(m_\pi c) N^{1/3}_p$, where $\Delta$ is a parameter such that $\Delta\approx 1$ ($\Delta <1$) corresponds to nuclear (supranuclear) densities when applied to ordinary nuclei, i.e. for $N_p/A\approx 1/2$, being $A=N_p+N_n$ the total nucleon number and $N_n$ the total number of neutrons. The distribution of $N_e=N_p$ degenerate electrons, $n_e$, subjected to the equilibrium condition determined by the constancy of their Fermi energy, $E^F_e=\mu_e-m_ec^2-eV=$constant, where $\mu_e=\sqrt{(c P^F_e)^2+m^2_e c^4}$ and $V$ are the chemical and Coulomb potential, is computed self-consistently from the electrostatic Poisson equation, $\nabla^2 V(r)=-4 \pi e (n_p-n_e)$, with boundary conditions of global neutrality. The electron number density is then given by $n_e=(P^F_e)^3/(3\pi^2 \hbar^3)=(e^2V^2+2m_e c^2 e V)^{3/2}/(3\pi^2\hbar^3)$. The distribution of neutrons is hence obtained from the constraint of $\beta$ equilibrium. From the proton and electron densities, we obtain the charge density distribution $\rho_{\rm ch}=e (n_p-n_e)$, which allows us to compute the potential $F$, and finally the magnetic field from Eqs.~(\ref{eq:Bgeneral}--\ref{eq:magcomponent}).

For a neutron star rotating with a period $P\approx 10$s we obtain

\begin{equation}
{\rm Core}:           
\left\{
\begin{array}{l}
B_r\sim|B_{\theta}|\sim 3\times 10^{-19}B_c\, , \\                       
B_{\rm core}=\sqrt{B_r^2+B_{\theta}^2}\approx 10^{-19}B_c\, ;
\end{array}
\right.
\end{equation}

\begin{equation}
{\rm Shell}:
\left\{
\begin{array}{l}
B_r\sim3\times 10^{-19}B_c \;,\; |B_{\theta}|\sim 10^{-1}B_c\, , \\
B_{\rm shell}=\sqrt{B_r^2+B_{\theta}^2}\approx 10^{-1}B_c\, .
\end{array}
\right.
\end{equation}

We can conclude from the above estimates that the magnetic field in the core is enough small to safely neglect its effect on the structure of the neutron star. We can also check the possible effects on the shell's structure. The magnetic, Coulomb, rotational and gravitational energy of the shell can be estimated respectively as
\begin{eqnarray}
\mathcal{E}_{\rm mag} &\approx& 0.446 \frac{\pi \hbar^2}{\alpha^{1/2} m_{\pi} c^2} \frac{N_p^{4/3}}{P^2}\, ,\\
\mathcal{E}_{\rm el} &\approx& 0.195 \frac{\pi^{1/2} m_{\pi} c^2}{\alpha^{1/2}} N_p^{2/3}\, ,\\
\mathcal{E}_{\rm rot} &\approx& 2 \frac{m_n \pi \hbar^2}{\alpha^{1/2} m_{\pi}^2 c^2}\frac{N_p^{4/3}}{P^2}\, ,\\
\mathcal{E}_{g} &\approx& -3 \frac{G m_{\pi} m_n^2 c}{\alpha^{1/2} \hbar} N_p^{1/3} A\, ,
\end{eqnarray}
where we have used as thickness of the shell, $\delta R_c \approx \hbar/(\sqrt{\alpha} m_{\pi} c)$, being $m_p$, $m_n$ and $m_\pi$ the neutron, proton and pion masses respectively, and $\alpha$ the fine structure constant. We therefore obtain
\begin{eqnarray}
\frac{\mathcal{E}_{\rm mag}}{|\mathcal{E}_{g}|} &\approx& 0.15\pi \left(\frac{m_{\rm Pl}}{m_n}\right)^2 \left(\frac{\hbar}{m_\pi c}\right)^2\frac{N_p/A}{(c P)^2}\approx 3.8\times 10^{-13},\\
\frac{\mathcal{E}_{\rm el}}{|\mathcal{E}_{g}|}&\approx& 0.06\pi^{1/2} \left(\frac{m_{\rm Pl}}{m_n}\right)^2 \frac{N^{1/3}_p}{A}\approx 0.05,\\
\frac{\mathcal{E}_{\rm rot}}{|\mathcal{E}_{g}|} &\approx& \frac{2\pi}{3} \left(\frac{m_n}{m_\pi}\right) \left(\frac{m_{\rm Pl}}{m_n}\right)^2 \left(\frac{\hbar}{m_\pi c}\right)^2\frac{N_p/A}{(c P)^2}\approx 1.2\times 10^{-11},
\end{eqnarray}
where $m_{\rm Pl}=(\hbar c/G)^{1/2}$ is the Planck's mass, and for the numerical estimates we have used a rotation period $P=10$~s, $N_p/A\approx 1/50$ and $A=10^{57}$.

We can see that both the rotational and magnetic energy are negligible corrections to the shell's energy for a rotation period $P=10$~s, being the main contributions owing to the gravitational and the electrostatic energy. 

It is clear that the above induced magnetic field in globally neutral neutron stars cannot be an explanation to the observed surface magnetic fields in pulsars since the induced magnetic field exists only in the interior up to the crust edge where global neutrality is reached, and therefore does not emerge up to the neutron star surface. The nature of the magnetic field observed in pulsars represents still a major issue in astrophysics and it is not the objective of the present work to try to answer such a question. The interior magnetic field in the neutron star can be larger than the one observed in its surface, however it is known that the effects of the magnetic field on the properties of nuclear matter at the high supranuclear densities present in the cores of neutron stars are expected to be appreciable only for extremely, and likely unrealizable, huge values $B\gtrsim 10^{18}$~G \citep[see, e.g.,][and references therein]{2012PhRvD..86l5032S,2012PhLB..707..163I,2013NuPhA.898...32D,2013PhRvC..88c5804D}. It implies that magnetic fields lower than these values do not have appreciable effects either on the nuclear equation of state or on the structure parameters of the neutron star \citep[see, also,]{1995A&A...301..757B,2000ApJ...537..351B}. It becomes clear that the effect of the low value of the magnetic field induced by electric field rotation in rotating globally neutral neutron stars, and of the possible interior magnetic field possibly present in the star's interior, can be safely neglected in the computation of the structure parameters, validating the treatment applied in this work. As we show below in the next section, more importantly are the innegable general relativistic effects affecting the radiation field near the surface of a rotating magnetic-dipole, i.e. the neutron star, and which can drastically modify the estimate of the surface magnetic field.

\section{Inference of pulsar's properties}\label{sec:3}

We turn now to the analysis of the consequences of using realistic general relativistic structure parameters on the inference of the magnetic field and efficiency of a pulsar in converting rotational energy into electromagnetic radiation. We focus here on the high-magnetic field pulsar class \citep[see][]{kaspi11} but our general qualitative results apply to all pulsars. 

As we have already stressed, the simplified picture of a point-like magnetic dipole has been traditionally applied as a model for pulsars of any rotation period and assuming fiducial values for the neutron star structure parameters. It is possible identify four major corrections that might be introduced to the model: 1) the existence of plasma magnetosphere instead of electrovacuum; 2) the dependence on the properties of the interior (equation of state) by the neutron star structure parameters such as mass, radius, and moment of inertia, with respect to the oversimplification lead by the use of fiducial values; 3) the effects due to the relativistic fast rotation, as measured by the fastness parameter, $\Omega R/c$; 4) the corrections measured by the compactness parameter $GM/(c^2R)$, introduced by the finiteness of the mass and size of the star. We discuss now each of these points.

The first correction depends upon the specific model of the pulsar's magnetosphere, which determines the electric potential developed above the neutron star surface and responsible for the acceleration of particles forming a wind that exerts a torque on the pulsar. However, starting from the classic work of \citet{1969ApJ...157..869G}, many competing models of the pulsar's magnetosphere have been proposed and they are still matter of debate in the literature. Therefore, we will not consider this issue in the present work.

Concerning the second point, we have shown in section \ref{sec:2} how the structure parameters depend on the neutron star theory and equation of state, nuclear fermion interactions strongly influence the mass-radius relation \citep[see, e.g.,]{2007PhR...442..109L}, hence all the derived pulsar parameters. Therefore different inferences of the magnetic field value can be obtained as a function of the neutron star mass, and nuclear equation of state.

The generalization of the Deutsch's results to the case of relativistic rotation ($\Omega \sim c/R$) was obtained by \citet{1992ApJ...401L..27B}. The radiation power in this case was expressed via a cumbersome integral which has to be solved numerically. The only exception is represented by the analytic expressions in the non-relativistic approximation, which leads to the Deutsch's solution, and in the ultra-relativistic approximation when $\Omega$ approaches $c/R$. The Maxwell equations are still solved there in flat Minkowski spacetime. This specific correction is expected to be important for millisecond pulsars. However, for the pulsar class discussed in this work, with rotation periods $P\sim 10$~s (hence, $\Omega R/c=2\pi R/(c P)\sim 10^{-5}$), such a correction is negligible and the solution in the slow rotation regime is sufficiently accurate. 

We focus now on the fourth correction. The exact solution of the exterior electromagnetic fields of a (slowly) rotating magnetic dipole aligned with the rotation axis in general relativity was first found by \citet{ginzburg64,ginzburg65}, see, also, \citet{1970Ap&SS...9..146A}. They solved the Einstein-Maxwell equations in the Schwarzschild background. The generalization to a general electromagnetic multipolar structure in a Schwarzschild metric was found by \citet{1970Ap&SS...9..146A}. The generalization of the Deutsch's solution to the general relativistic case in the slow rotation regime, and for a general misaligned dipole, was obtained in analytic form in the \emph{near zone} ($r \ll c/\Omega=1/k=\lambda/2\pi$) by \citet{2001MNRAS.322..723R,2003MNRAS.338..816R} and, for the wave zone by \citet{2004MNRAS.352.1161R}. In the latter, the radiation power of the dipole was computed as
\begin{equation}\label{eq:dipoleGR}
P^{\rm G.R.}_{\rm dip}=-\frac{2}{3}\frac{\mu^2_\bot \Omega^4}{c^3} \left(\frac{f}{N^2}\right)^2\, ,
\end{equation}
where $f$ and $N$ are the general relativistic corrections
\begin{eqnarray}\label{eq:fN}
f &=&-\frac{3}{8}\left(\frac{R}{M_0}\right)^3\left[\ln(N^2)+\frac{2 M_0}{R}\left(1+\frac{M_0}{R}\right)\right]\, ,\\
N &=&\sqrt{1-\frac{2 M_0}{R}}\, ,
\end{eqnarray}
being $M_0$ the mass of the non-rotating configurations. Now by equating the rotational energy loss to the above electromagnetic radiation power it is possible to obtain the formula of the surface magnetic field analogous to Eq.~(\ref{eq:Bmax}) but with general relativistic corrections:
\begin{equation}\label{eq:BGR}
B\sin\chi=\frac{N^2}{f} \left(\frac{3 c^3}{8 \pi^2} \frac{I}{R^6} P \dot{P} \right)^{1/2}\, .
\end{equation}

\begin{figure}[!hbtp]
\plotone{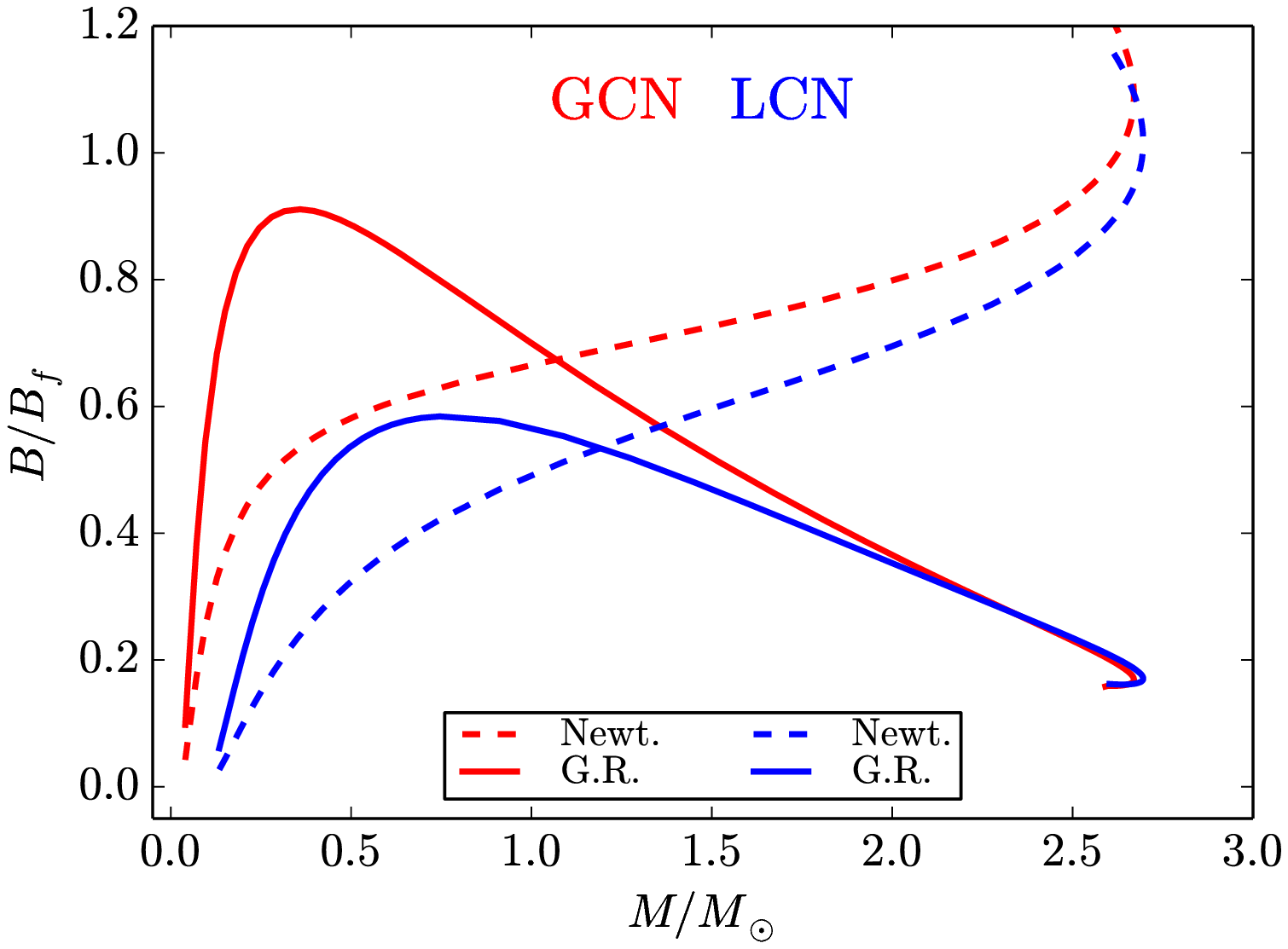}
\caption{Ratio of the magnetic field given by the Newtonian formula (\ref{eq:Bmax}) and the general relativistic one (\ref{eq:BGR}) to the fiducial value given by Eq.~(\ref{eq:BmaxNS}). We have used here the realistic mass-radius relations of globally and locally neutral static neutron stars of this work and an inclination angle $\chi=\pi/2$.}\label{fig:BvsBf}
\end{figure}

In Fig.~\ref{fig:BvsBf} we have plotted the ratio of the magnetic field obtained via the Newtonian formula (\ref{eq:Bmax}) and the general relativistic formula (\ref{eq:BGR}) to the fiducial value obtained with (\ref{eq:BmaxNS}), for the realistic mass-radius relations of globally and locally neutral neutron stars used in this work.

We can see from this figure that, in the Newtonian case, the inferred magnetic field increases with increasing neutron star mass. Therefore, the configurations of maximum and minimum mass give us respectively, in such a case, upper and lower limits to the magnetic field. It is worth noticing how the general relativistic formula gives us a magnetic field lower than the Newtonian counterpart, for $M/M_{\odot}\gtrsim 1.1$ and $M/M_{\odot}\gtrsim1.2$ for the globally and locally neutral configurations respectively. In addition, we find a markedly different and interesting behavior. First, the magnetic field is extremely close, at very low masses, with the Newtonian value, as expected; then, it deviates and reaches a maximum value for some value of the mass, and then decreases for increasing masses. The magnetic field inferred from globally and locally neutral configurations coincides for large masses close to the critical mass value, as it should be expected since in those massive configurations the structure parameters are dominated by the neutron star core, with very little role of the crust. We are here using the parameters of the static configurations. This is a good approximation for this family of pulsars since their rotation periods are well far the millisecond region, where deviations from spherical symmetry are expected. This can be seen in Fig.~\ref{fig:MtotvsRtot}, where the sequence of constant rotation period $P=10$~s essentially overlaps the static mass-radius relation.

\begin{figure*}[!htbp]
\plottwo{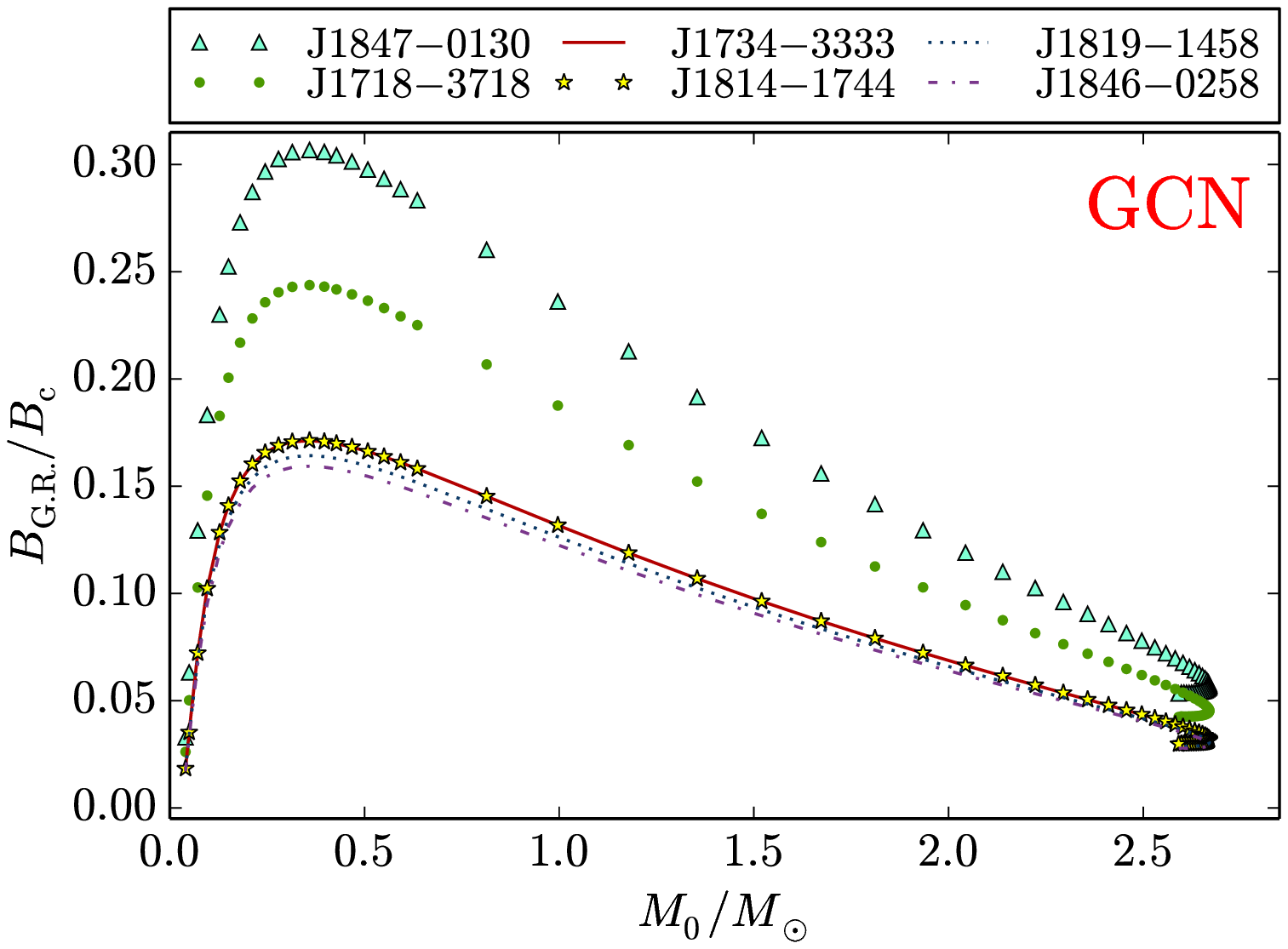}{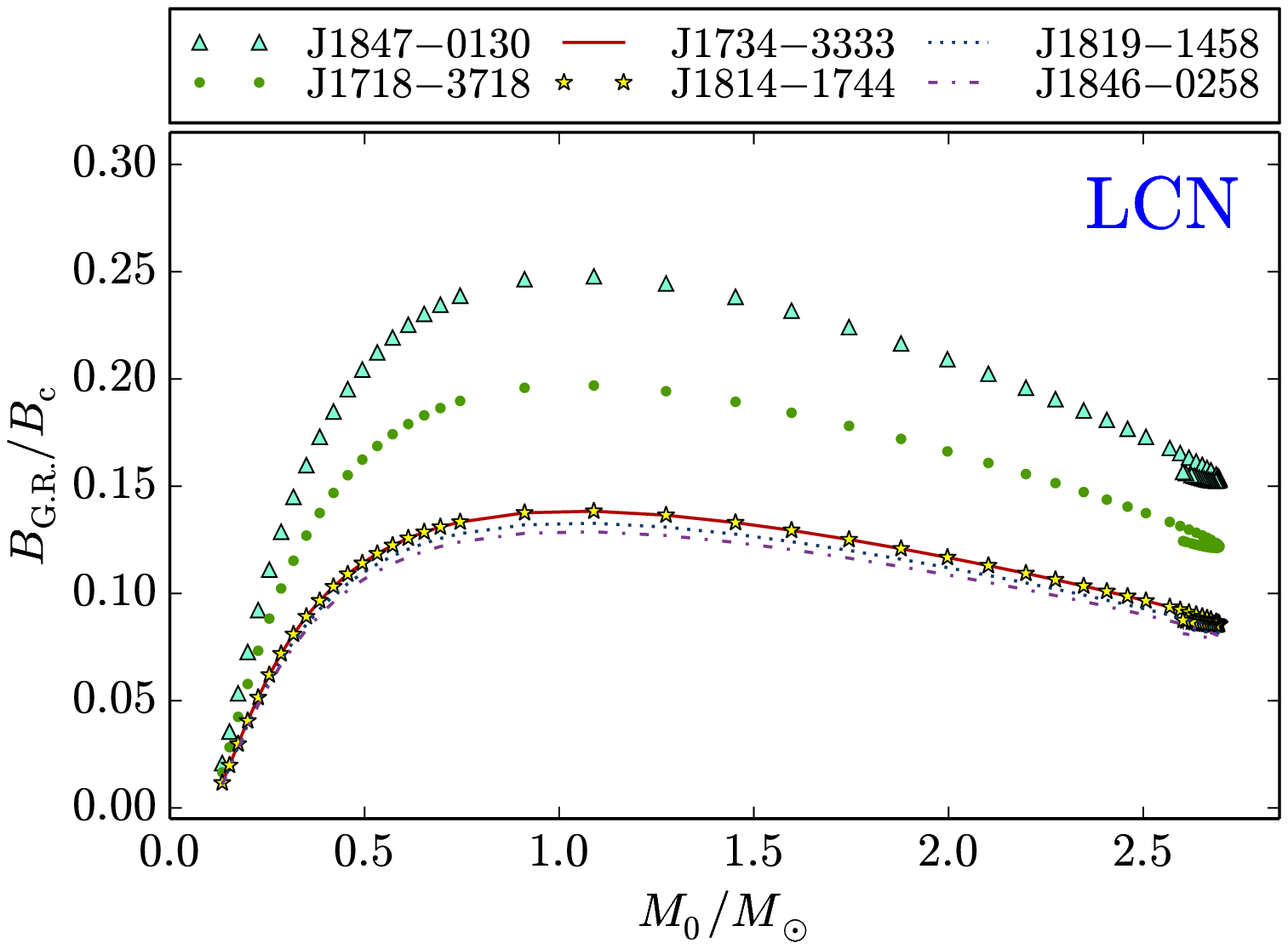}
\caption{Magnetic field $B_{\rm G.R.}$ obtained from the general relativistic magneto-dipole formula (\ref{eq:BGR}), in units of critical magnetic field $B_{\rm c}$, as function of the mass (in solar masses) for static neutron stars in the global (left panel) and local (right panel) charge neutrality cases. }\label{fig:magneticstatic}
\end{figure*}

In Fig.~\ref{fig:magneticstatic} we plotted our theoretical prediction for magnetic fields of the pulsars of Table \ref{tab:H-B_pulsar} as a function of the neutron star mass, using the general relativistic formula (\ref{eq:BGR}).

We find that, both in global and local neutrality case, the assumed high-B pulsars have inferred magnetic fields lower than the critical value for the entire range of neutron star masses.

\begin{figure*}[!hbtp]
\plottwo{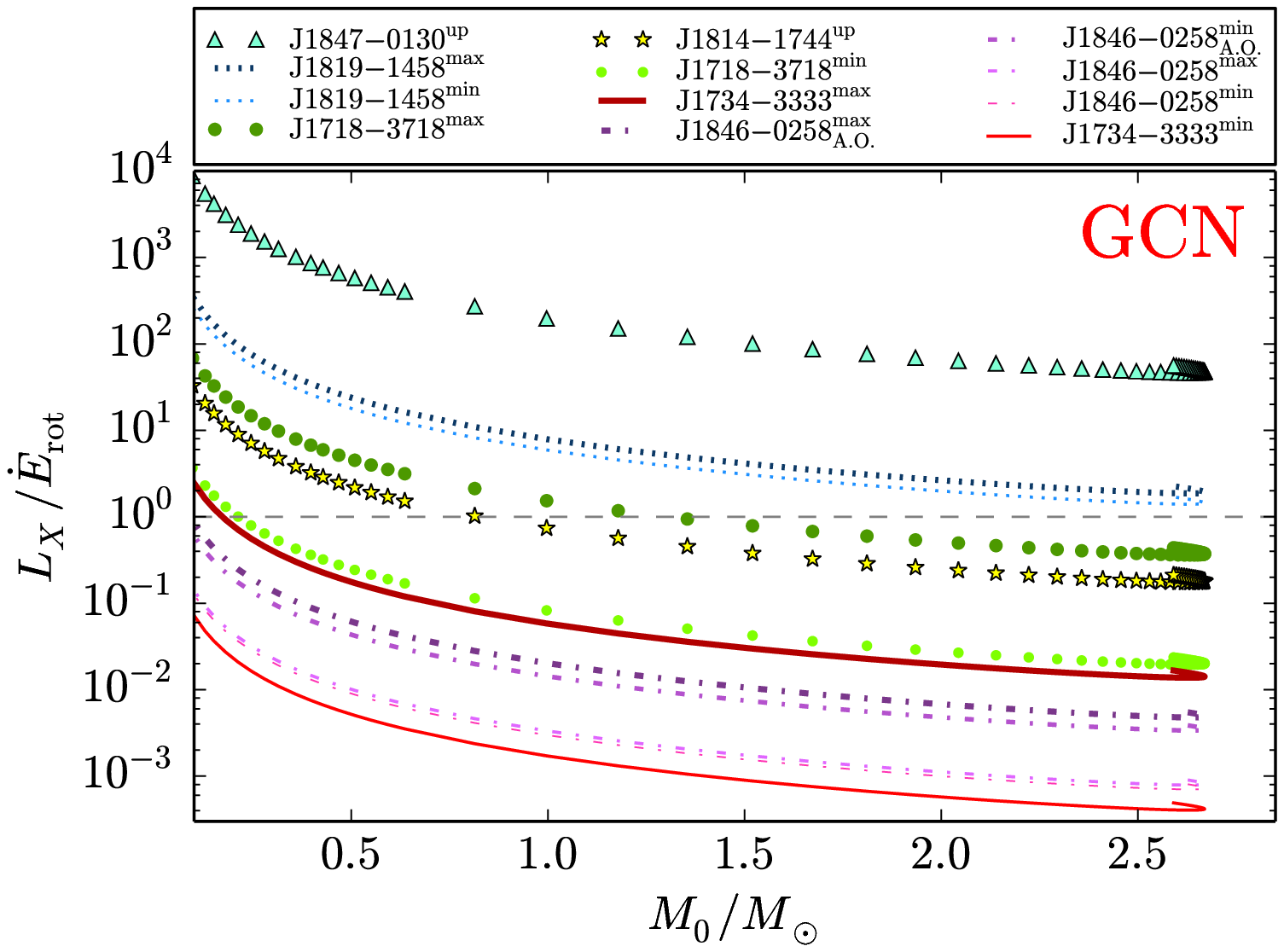}{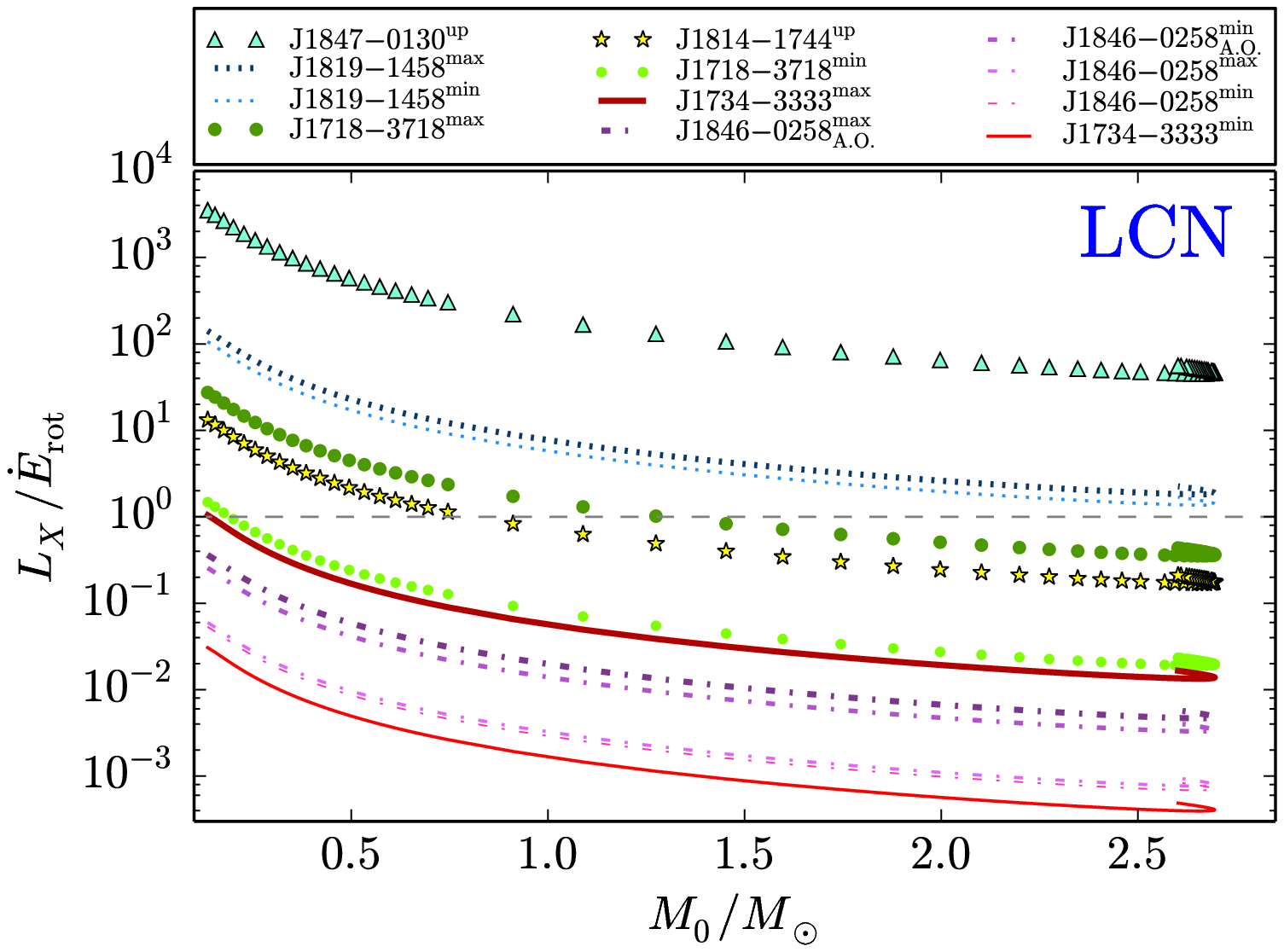}
\caption{Ratio between the observed X-ray luminosity $L_X$ and the loss of rotational energy $\dot{E}_{\rm rot}$ versus total mass of the rotating neutron star, in units of $M_{\odot}$. Are drawn the high-B pulsar from the work by \citet{kaspi11} for which a magnetic field higher than the critical field $B_{\rm c}$ is inferred, once the fiducial value for the moment of inertia $I=10^{45}$~g~cm$^{2}$ is taken into account (see Table \ref{tab:H-B_pulsar}). Pulsars with luminosity  $L_X$ defined by an upper limit are labeled with ``up'',  for pulsars with luminosity  $L_X$ not well established we have assumed the existent lower limits (label ``min'') and upper limits (label ``max'') on it.  The values for the pulsar PSR J1846-0258 are dived in prior the $2006$ outburst and after the $2006$ outburst (label ``A.O.''). Left plot: global charge neutrality. Right plot: local charge neutrality. The magnetic fields shown are referred to  the high-magnetic field pulsars of Table \ref{tab:H-B_pulsar}.}\label{fig:LxoverEdot}
\end{figure*}

Concerning the efficiency of pulsars in converting rotational energy into electromagnetic radiation, we show in Fig.~\ref{fig:LxoverEdot} the X-ray luminosity to rotation energy loss ratio, $L_X/\dot{E}_{\rm rot}$, as a function of the neutron star mass, for both global and local charge neutrality. For the sake of comparison, we also present in Table \ref{tab:H-B_pulsar} the ratio $L_X/\dot{E}_{\rm rot}^f$, where $\dot{E}_{\rm rot}^f$ is the rotational energy loss as obtained from fiducial neutron star parameters given by Eq.~(\ref{eq:Edotf}).

We find that for both globally and locally neutral neutron stars we have $L_X<\dot{E}_{\rm rot}$: 1) in PSR J$1718$--$3718$ for $M_0\gtrsim 1.25~M_\odot$ and for the entire range of masses adopting, respectively, the observational upper or lower limits on $L_X$; 2) in PSR J$1814$--$1744$ for $M_0\gtrsim 0.8~M_\odot$ using the upper limit on $L_X$; 3) for the rest of the objects in the entire range of stable masses. 

The only exception to the above result are PSR J1847--0130 and PSR J1819--1458, for which no range of masses with $L_X<\dot{E}_{\rm rot}$ was obtained. However, for PSR J$1847$--$0130$ we have only an upper limit for $L_X$, so there is still room for solutions with $L_X<\dot{E}_{\rm rot}$ if future observations lead to an observed value lower than the present upper limit. In this line, the only object with $L_X>\dot{E}_{\rm rot}$ for any mass is PSR J$1819$--$1458$. For this particular object there is still the possibility of being a rotation powered neutron star since the currently used value of the distance to the source, 3.6 kpc, inferred from its dispersion measure, is poorly accurate with a considerable uncertainty of at least 25\% \citep[see][for details]{mclaughlin07}. Indeed, a distance to the source 25\% shorter than the above value would imply $L_X<\dot{E}_{\rm rot}$ for this object in the mass range $M_0\gtrsim 0.6~M_\odot$.

We notice that the efficiency obtained via fiducial parameters, $L_X/\dot{E}_{\rm rot}^f$, is larger than the actual value obtained with the realistic neutron star structure in the entire range of stable masses; see Table~\ref{tab:H-B_pulsar} and Fig.~\ref{fig:LxoverEdot}.

It is also worth to mention that the rotation energy loss (\ref{eq:Edot}) depends on the neutron star structure only through the moment of inertia, whose quantitative value can be different for different nuclear EOS and/or owing to an improved value accounting for deviations from the spherical geometry, for instance considering  a third-order series expansion in $\Omega$. However, the latter effect is negligible for this specific case ($P\approx 4.3$~s), see for instance Fig.~5 in \citep{benhar05}, where no deviations of $I$ from its spherical value appear for such long rotation periods.

\section{Concluding remarks}\label{sec:4}

We explored the consequences of a realistic model for neutron stars on the inference of the astrophysical observables of pulsars. We showed in particular that: 

1) The magnetic field is overestimated when fiducial parameters are adopted, independently on the use of either the Newtonian or the general relativistic radiation formula of the rotating magnetic dipole; see Fig.~\ref{fig:BvsBf}.

2) The use of the Newtonian formula (\ref{eq:Bmax}) can overestimate the surface magnetic field of up to one order of magnitude with respect to the general relativistic one given by Eq.~(\ref{eq:BGR}). We applied these considerations to the specific case of the high-magnetic field pulsar class, for which overcritical magnetic fields have been obtained in the literature with the use of fiducial neutron star parameters within the Newtonian rotating magnetic dipole model, i.e. estimating the magnetic field through Eq.~(\ref{eq:BmaxNS}). We found that, instead, the magnetic field inferred for these pulsars turn to be undercritical for any values of the neutron star mass; see Fig.~\ref{fig:magneticstatic}.

3) The nontrivial dependence of the inferred magnetic field on the neutron star mass, in addition to the dependence on $P$ and $\dot{P}$, namely $B=B(I(M_0),R(M_0),P,\dot{P})$, leads to the impossibility of accommodating the pulsars in a typical $\dot{P}-P$ diagram together with a priori fixed values of the magnetic field; see Fig.~\ref{fig:magneticstatic}.

4) We computed the range of neutron star masses for which the X-ray luminosity of these pulsars can be well explained via the loss of rotational energy o the neutron star and therefore falling into the family of ordinary rotation-powered pulsars. The only possible exceptions were found to be PSR J1847--0130 and PSR J1819--1458, which however, as we argued, still present observational uncertainties in the determination of their distances and/or luminosities that leave room for a possible explanation in terms of spindown power. We also showed that the efficiency of the pulsar, $L_X/\dot{E}_{\rm rot}$, is overestimated if computed with neutron star fiducial parameters.

5) We discussed the possible effects of different nuclear models as well as the improved values of the moment of inertia given by further expansion orders of the slow rotation approximation or full numerical integration of the equilibrium equations in the rotating case. However, the former effect appears to be negligible for the long rotation periods, $P\sim 10$~s, of the high-magnetic field pulsars \citep[see e.g.~Fig.~5 in][]{benhar05}. We have also given estimates of the magnetic field induced by rotation of the interior charge distribution in neutron stars satisfying the condition of global, but not local, charge neutrality. We have shown that for the case of these long rotational periods the effects of the magnetic field, both in the core and in the core-crust transition surface of these configurations are, in first approximation, negligible.

It is worth to underline that the validity of the results of this work very likely apply also to different nuclear EOS consistent with the current observational constraints, as suggested by the high value of the recently measured mass of PSR J0348+0432, $M=2.01 \pm 0.04 M_\odot$ \citep{antoniadis13}. Such a high value favors stiff nuclear EOS, as the one used here based on relativistic nuclear mean field theory \'a la \cite{boguta77b}, which lead to a critical mass of the neutron star higher than the above value.

\emph{Acknowledgements.--} It is a pleasure to thank K.~Boshkayev for helpful discussions on the simplified model to estimate the rotationally induced magnetic field in the core and in the core-crust interface of the neutron star. We would like to thank the referee for the very constructive comments which led to an improvement of the presentation of our results. R.B. and J.A.R. acknowledge the support by the International Cooperation Program CAPES-ICRANet financed by CAPES -- Brazilian Federal Agency for Support and Evaluation of Graduate Education within the Ministry of Education of Brazil.

\bibliographystyle{apj}
\bibliography{biblio}

\end{document}